\documentclass[sigconf]{acmart}

\copyrightyear{2026}
\acmYear{2026}
\setcopyright{cc}
\setcctype{by}
\acmConference[MSR '26]{23rd International Conference on Mining Software Repositories}{April 13--14, 2026}{Rio de Janeiro, Brazil}
\acmBooktitle{23rd International Conference on Mining Software Repositories (MSR '26), April 13--14, 2026, Rio de Janeiro, Brazil}
\acmPrice{}
\acmDOI{10.1145/3793302.3793596}
\acmISBN{979-8-4007-2474-9/2026/04}

\usepackage{subcaption}
\usepackage{multirow}

\begin{document}

\title{On the Adoption of AI Coding Agents in Open-source Android and iOS Development}

\author{Muhammad Ahmad Khan}
\email{25030004@lums.edu.pk}
\affiliation{%
  \institution{Lahore University of Management Sciences}
  \city{Lahore}
  \country{Pakistan}
}

\author{Hasnain Ali}
\email{25280088@lums.edu.pk}
\affiliation{%
  \institution{Lahore University of Management Sciences}
  \city{Lahore}
  \country{Pakistan}
}

\author{Muneeb Rana}
\email{muneebrana625@gmail.com}
\affiliation{%
  \institution{Xtra App Studios}
  \city{Multan}
  \country{Pakistan}
}

\author{Muhammad Saqib Ilyas}
\email{saqibm@lums.edu.pk}
\affiliation{%
  \institution{Lahore University of Management Sciences}
  \city{Lahore}
  \country{Pakistan}
}

\author{Abdul Ali Bangash}
\email{abdulali@lums.edu.pk}
\affiliation{%
  \institution{Lahore University of Management Sciences}
  \city{Lahore}
  \country{Pakistan}
}

\begin{abstract}
AI coding agents are increasingly contributing to software development, yet their impact on mobile development has received little empirical attention. In this paper, we present the first category-level empirical study of agent-generated code in open-source mobile app projects. We analyzed PR acceptance behaviors across mobile platforms, agents, and task categories using 2,901 AI-authored pull requests (PRs) in $193$ verified Android and iOS open-source GitHub repositories in the AIDev dataset. We find that Android projects have received 2x more AI-authored PRs and have achieved higher PR acceptance rate ($71$\%) than iOS ($63$\%), with significant agent-level variation on Android. Across task categories, PRs with routine tasks (\textit{feature}, \textit{fix}, and \textit{ui}) achieve the highest acceptance, while structural changes like \textit{refactor} and \textit{build} achieve lower success and longer resolution times. Furthermore, our evolution analysis shows improvement in PR resolution time on Android through mid-2025 before it declined again. Our findings offer the first evidence-based characterization of AI agents effects on OSS mobile projects and establish empirical baselines for evaluating agent-generated contributions to design platform aware agentic systems.
\end{abstract}

\maketitle

\section{Introduction}

AI coding agents like Codex, and Copilot are increasingly contributing to software development, doing code changes and fixes across diverse projects. Large datasets such as AIDev \cite{li2025aidev}, which contains over 932k AI-authored pull requests (PRs) across 116k open-source GitHub repositories, have now enabled empirical analysis of these activities at scale. However, the behavior of such agents in \textit{mobile app development} has received little empirical attention.

Mobile development involves complex settings of design, logic and hardware configurations. Mobile platforms, such as Android and iOS, differ in build systems, language ecosystems, project structures, and CI workflows \cite{6681334,androidioszazz}, and these differences can shape the workflow and review process of agent-authored PRs. Despite the prominence of mobile platforms, no prior empirical work has examined AI coding agents in mobile app projects.

To address this gap, we analyzed $2,901$ AI-authored PRs in $193$ verified Android and iOS open-source GitHub repositories from the AIDev dataset \cite{li2025aidev}. We classified PRs into $13$ empirically derived task categories using \textit{Open-card sorting} with GPT-5, which authors refined and manually validated.

We investigate the following research questions:
\begin{itemize}
    \item \textbf{RQ1:} How do AI coding agents differ in PR acceptance rate across mobile platforms?
    \item \textbf{RQ2:} How does PR acceptance rate vary across categories?
    \item \textbf{RQ3:} How does PR resolution time evolve across PR categories on mobile platforms over time?
\end{itemize}

Our work provides actionable insights for maintainers to optimize review workflows. By identifying key platform differences, specifically Android's sensitivity to agent choice versus iOS's uniform caution, we offer a blueprint for building the next generation of platform-aware agentic systems.

\textbf{Replication Kit:} To facilitate replication and further studies, we provide the data and code used in our study.\footnote{https://zenodo.org/records/17874506}

\section{Dataset}

We use the \textbf{AIDev dataset}~\cite{li2025aidev}, which contains 932k AI-authored pull requests (PRs) across 116k open-source GitHub repositories. Five agents (\textit{Codex, Devin, Copilot, Cursor, and Claude}) generated these PRs between May-Jul 2025.

From this corpus, we identify mobile app projects and retrieve all AI-authored PRs from those repositories. Since AIDev window is small (May-Jul 2025), we extended this dataset by mining additional AI-authored PRs (Aug–Nov 2025) from the selected mobile projects using the GitHub API \cite{github_api_docs}.

\section{Methodology}
\label{sec:method}

\begin{figure}[!htb]
    \centering
    \includegraphics[width=\linewidth]{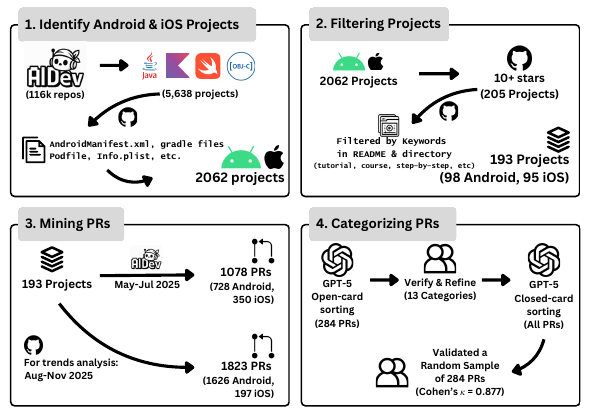}
    \caption{Overview of our methodology.}
    \Description{Overview of our methodology.}
    \label{fig:methodology}
\end{figure}

We used a four-step process (Figure~\ref{fig:methodology}) to filter mobile projects, fetch AI-authored PRs, and categorize PRs by task.

\subsection{Identification of Android \& iOS Projects}

From the AIDev dataset, we selected repositories in Java, Kotlin, Swift, or Objective-C to target native Android and iOS development. We chose native Android and iOS projects to study AI agent behavior under platform-specific development constraints which may also be followed by cross-platform frameworks. We ensure domain relevance through rule-based checks using GitHub APIs. \textbf{For Android}, we required the presence of \texttt{AndroidManifest.xml}, Gradle files, and Android-specific directory structures of the Github repository. \textbf{For iOS}, we verified the presence of Xcode artifacts such as \texttt{Info.plist}, \texttt{.xcodeproj}, \texttt{Podfile}, and structured Swift layouts in the Github repository.

\subsection{Project Filtration}

After identification of Android and iOS projects, we filter out projects with less than 10 stars. We further filter out repositories to remove tutorials and samples by identifying indicators (e.g., ``tutorial'', ``step-by-step'', ``course'') in topics, README text, or directory structure of the Github repositories. This exclusion left us with 193 projects (98 Android and 95 iOS) for analysis.

\subsection{Mining AI-authored PRs}

From the 193 filtered repositories, we extracted $1,078$ AI-authored PRs ($728$ Android, $350$ iOS) from the AIDev dataset (May–Jul 2025). For temporal analysis in RQ3, we extend the AIDev dataset by mining all AI-authored PRs from Aug to Nov 2025 using the same GitHub search queries used for the AIDev dataset~\cite{li2025aidev}. This process added $1,823$ additional PRs ($1,626$ Android, $197$ iOS). Please note that we cover the complete space of OSS repositories in the AIDev dataset, so we retain the natural Android–iOS imbalance, as it reflects actual differences in agent activity across both ecosystems.

\begin{table}[h]
\centering
\caption{Summary of extracted data.}
\vspace{-3mm}
\begin{tabular}{lcc}
\hline
\textbf{Item} & \textbf{Android} & \textbf{iOS} \\
\hline
Language-filtered repositories & 4,532 & 1,106 \\
Filtered projects ($\ge$10 stars) & 107 & 98 \\
Final repositories (Non-tutorial) & 98 & 95 \\
AI-authored PRs - AIDev (May-Jul 2025) & 728 & 350 \\
AI-authored PRs - Extended (Aug-Nov 2025) & 1,626 & 197 \\
\hline
\end{tabular}
\label{tab:summary_data}
\end{table}

\subsection{PR Categorization}

Because AIDev limits its task annotations to high-star (100+) repositories, we categorize all PRs in our dataset (including extended set of Aug–Nov 2025). To categorize PRs into their relevant task categories, we first randomly sample $284$ PRs with a $95\%$ confidence interval and $5\%$ error margin from $1,078$ PRs.
Then, we perform open-card sorting by providing the PR-titles to GPT-5, which returned the probable PR categories at its default temperature~\cite{zhao2024empirical}.
Two of the paper authors then refine and finalize the GPT-5 suggested categories through a negotiated-agreement \cite{Braun01012006}.
Finally, to cateogrize all the PRs, we provide GPT-5 the finalized categories and all the PRs titles.
To assess reliability, the two authors independently validated a random sample of $284$ PRs (with $95\%$ confidence interval, $5\%$ error margin) and achieved a strong inter-rater agreement (Cohen’s $\kappa = 0.877$).
Our finalized $13$ categories were the following: 
 \textit{feature}, \textit{fix}, \textit{refactor}, \textit{build}, \textit{chore}, \textit{performance}, \textit{style}, \textit{test}, \textit{docs}, \textit{operations}, \textit{ui}, \textit{localization}, and \textit{other}. 
 
\subsection{Statistical Analysis}

Each PR includes \texttt{created\_at}, \texttt{closed\_at}, and \texttt{merged\_at} timestamps. We computed PR resolution time as \texttt{merged\_at – created\_at}. Since some agents have only a few PRs in certain categories, the raw PR acceptance rates become sensitive to small samples. To address this issue, we applied Bayesian add-$\alpha$ smoothing \cite{jurafsky2000slp}. 

To answer our RQs, we used non-parametric tests (Mann–Whitney~U, Chi-Square, and Kruskal–Wallis) with Holm correction to compare distributions across mobile platforms, agents, and categories, enabling robust analysis under non-normal data.

\begin{figure}[t]
    \centering
    \includegraphics[width=\linewidth]{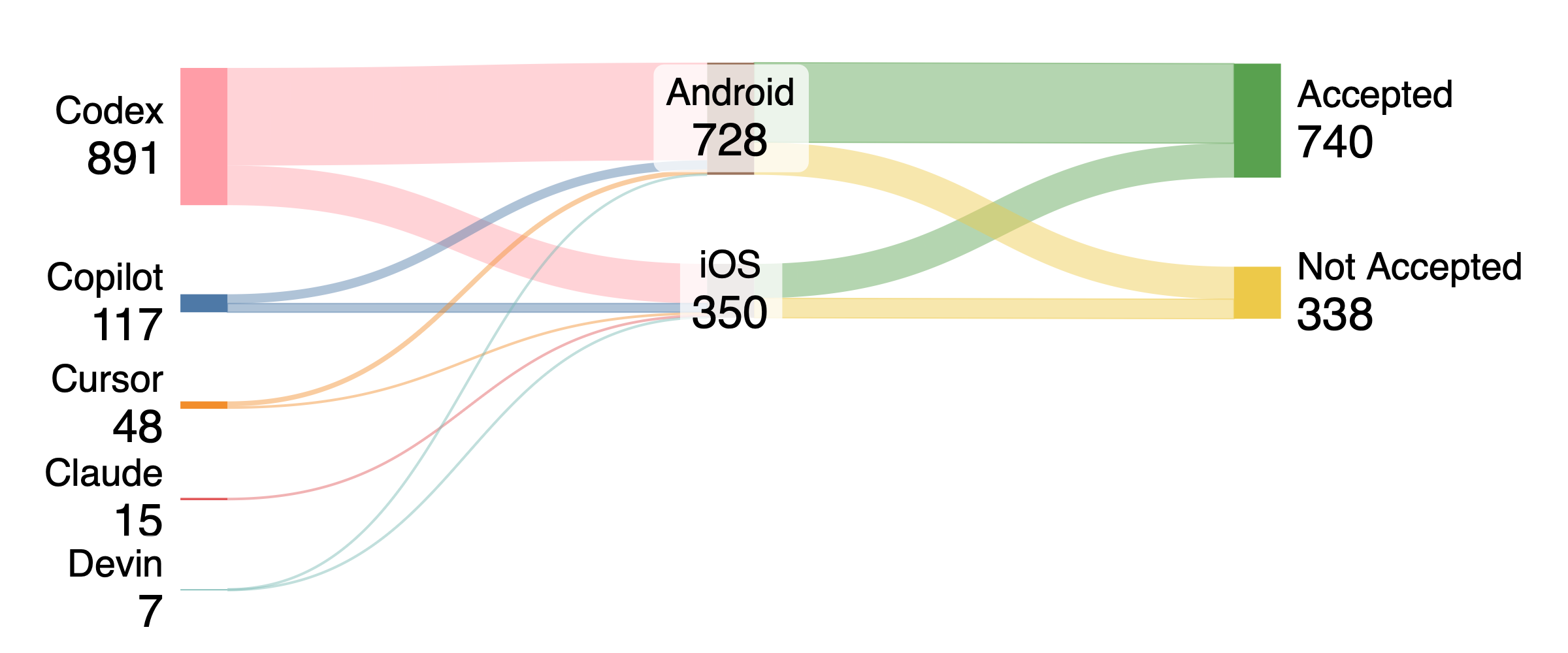}
    \vspace{-6mm}
    \caption{Agents' contributions across mobile platforms.}
    \Description{Agents' contributions across mobile platforms.}
    \label{fig:rq1}
\end{figure}

\begin{figure*}[ht]
    \centering
    \includegraphics[width=\textwidth]{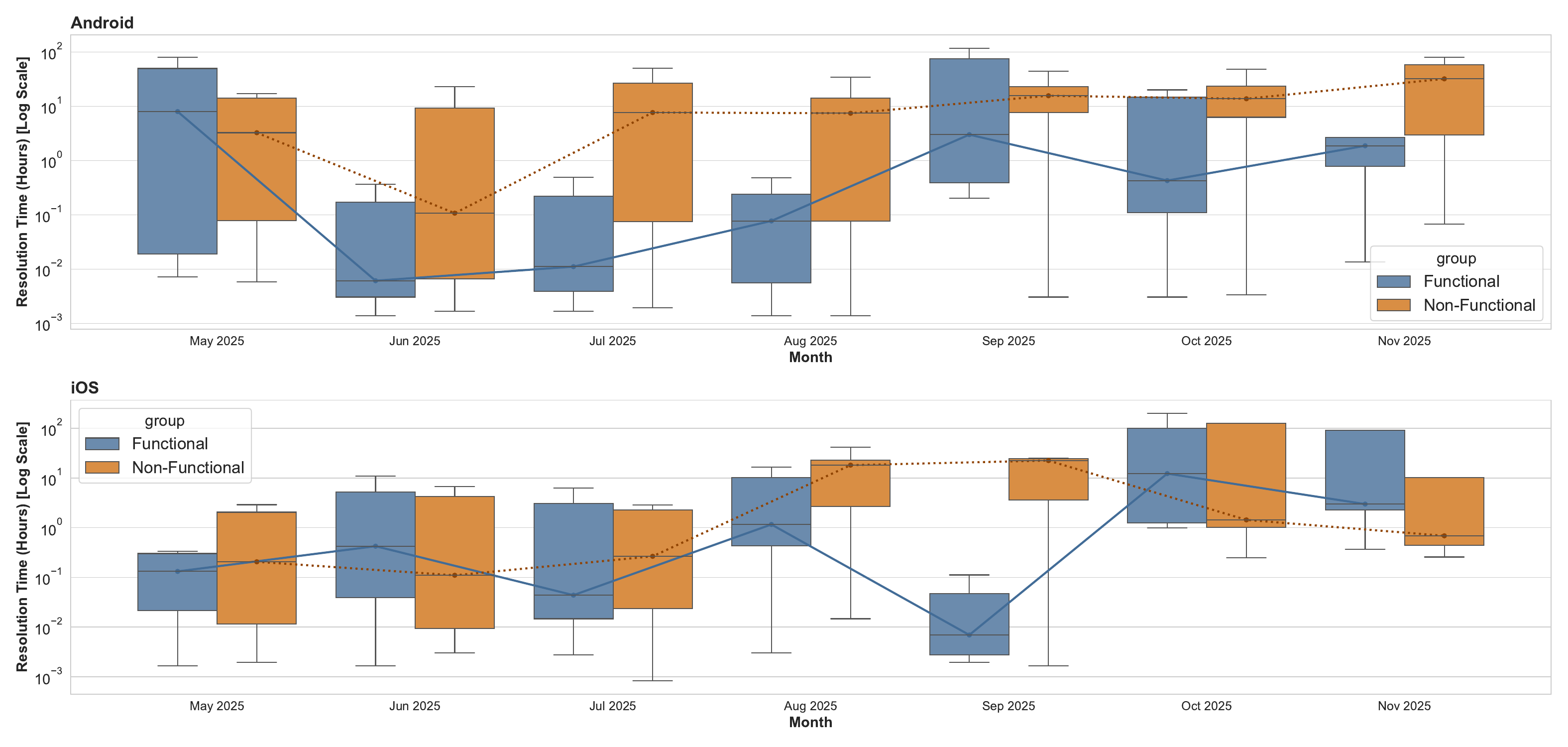}
    \vspace{-3em}
    \caption{PR resolution time trends across task groups (lower values = better) [log-scaled].}
    \Description{PR resolution time trends across task groups (lower values = better) [log-scaled].}
    \label{fig:resolution_category}
\end{figure*}
\section{Results}
\label{sec:results}

\subsection{RQ1: PR Acceptance Rate Across Mobile Platforms and Agents}

We analyze PR success across mobile platforms and agents using PR acceptance rates. Figure~\ref{fig:rq1} shows the overall agent's contributions landscape across mobile platforms. Android projects show a higher PR acceptance rate ($71.0\%$) than iOS ($63.7\%$), as Mann--Whitney~U test confirmed a statistical difference between the two ($p < 0.05$). Lower PR acceptance rate on iOS could be due to strict design guidelines and development environment \cite{androidioszazz}.

We further confirm using a Chi-square test that Agent-level performance vary significantly on Android ($p < 0.05$), but not on iOS ($p > 0.05$). 
On Android, maintainers accept \textit{Codex} PRs significantly more often ($76.8\%$) than \textit{Copilot} ($28.0\%$) and \textit{Cursor} ($42.3\%$). While on iOS, agents display relatively uniform PR acceptance rates and remain within a narrow ($51-79\%$) range.\\

\noindent\fbox{%
  \begin{minipage}{\dimexpr\linewidth-2\fboxsep-2\fboxrule\relax}
    \small
    \textbf{RQ1 Summary:} Android projects accept more AI-authored PRs than iOS, despite receiving a higher frequency of such PRs. Agent-level differences appear only on Android, where certain agents like Codex consistently achieve higher PR acceptance rate, while iOS shows minimal variation across agents.
    
    \textbf{Implication:} For developers and tool builders, our results indicate that agent choice matters more at the Android platform. Moreover, the consistently low and uniform PR acceptance rates on iOS require a more cautious stance toward AI-generated contributions.
  \end{minipage}%
}

\subsection{RQ2: PR Acceptance Rate Across Categories}

We analyze PR acceptance rates across 13 PR categories for the top three most contributing agents on each platform. 

On Android, Chi-square test confirm that categories are significantly different from each other in terms of PR acceptance rate ($p < 0.05$). Categories such as \textit{localization} (100\%), \textit{ui} ($88\%$), and \textit{fix} ($75\%$) have the highest PR acceptance rate. Pair-wise Fisher's exact post-hoc test shows that \textit{localization} PRs ($100\%$ acceptance) differ significantly from several low-performing categories, including \textit{refactor}, \textit{feature}, and \textit{build} ($p < 0.05$).

On iOS, Chi-square test shows no significant category-level variation ($p > 0.05$), and the post-hoc test also remains insignificant, indicating broadly uniform acceptance behavior across categories.\\

\noindent\fbox{%
\parbox{\linewidth}{%
\small
\textbf{RQ2 Summary:} Android projects show clear differences in PR acceptance rate across categories. Routine tasks such as \textit{localization}, \textit{ui}, and \textit{fix} achieve a higher PR acceptance rate than structural updates such as \textit{feature}, \textit{refactor}, and \textit{build}. While, iOS projects show largely uniform acceptance behavior across categories.

\textbf{Implication:} Our results highlight which PR task categories are most compatible with current AI agents, allowing Android teams to focus automation on categories with higher PR acceptance while allocating additional review effort to structural changes. Since iOS shows minimal category variation, such task-specific optimization may offer limited gains there.
}}

\subsection{RQ3: PR Resolution Time Trends Across Task Categories on Mobile Platforms}

Platform-level Mann–Whitney~U test shows that AI-authored PRs significantly vary across both mobile platforms ($p<0.05$). AI-authored PRs on iOS resolve $18\times$ faster than the PRs on Android.

A Kruskal–Wallis test across agents confirms that resolution time differs on both platforms (Android: $p<0.05$ and iOS: $p<0.05$). On Android, post-hoc Dunn's test shows significant differences between \textit{Codex} and \textit{Claude} ($p<0.05$), while other pairwise differences are insignificant. \textit{Codex} resolves PRs $3\times$ faster than \textit{Claude}.
On iOS, \textit{Devin} and \textit{Codex} differ significantly than all other agents (all $p<0.05$). Although \textit{Devin} appears fastest, its small sample size ($n=7$) prevents reliable inference; \textit{Codex} remains the fastest agent on both platforms.

Figure~\ref{fig:resolution_category} plots monthly PR resolution times. To better understand the trend, we divide the categories in \textbf{functional} \textit{(feature, ui, localization, fix)} and \textbf{non-functional} \textit{(refactor, build, chore, performance, style, test, docs, operations)} groups. On Android, Kruskal–Wallis test across months shows that functional PRs differ significantly than non-functional PRs ($p<0.05$). Functional PRs resolve $400\times$ faster than non-functional ones overall. PR resolution times improve from May to Aug, with distributions shifting downward, but increase again in later months, indicating a partial reversal of earlier efficiency gains. Post-hoc Dunn's test confirms this dynamic pattern, showing significant differences for both functional ($p<0.05$) and non-functional PRs ($p<0.05$).

On iOS, Kruskal–Wallis test across months shows that functional PRs also differ significantly than non-functional PRs ($p<0.05$). Functional PRs resolve $7\times$ faster than non-functional ones. Post-hoc Dunn's test shows significant month-to-month differences for both groups. However, the distributions show less systematic improvement, with higher and overlapping PR resolution times persisting across months. Overall, Android exhibits pronounced but unstable temporal improvements, whereas iOS shows statistically detectable yet visually modest temporal variation.\\

\noindent\fbox{%
\parbox{\linewidth}{%
\small
\textbf{RQ3 Summary:}
Overall, AI-authored PRs resolve faster on iOS than on Android, while Android shows stronger but unstable temporal dynamics. \textit{Codex} PRs are consistently resolved faster than other agents on both platforms. Category effects are pronounced on Android, where functional PRs resolve much faster than non-functional ones, whereas iOS shows weaker category separation. Temporally, Android improves through mid-2025 before declining again, while iOS exhibits statistically significant but visually modest month-to-month variation.

\textbf{Implication:}
PR resolution efficiency depends on mobile platform, PR task category, and agent choice. Android shows stronger but less stable efficiency gains that vary by PR task category, while iOS maintains faster yet more stable resolution behavior over time. These findings indicate that temporal improvements in AI-assisted development are shaped by platform-specific review dynamics.
}}

\section{Threats to Validity}

We used PR acceptance rate and PR resolution time as indicators for contribution outcomes, but these metrics do not capture code quality, review depth, or long-term code reliability. We derive the category taxonomy from PR titles using GPT-5 with default temperature that may misrepresent tasks with mixed or ambiguous intent, even with expert validation. The dataset reflects active open-source Android and iOS projects on GitHub, and results may differ in settings with other review cultures or CI practices. Some agents and categories contain limited samples, which may affect statistical robustness despite smoothing and non-parametric testing with correction. These limitations outline the scope of our findings and indicate where future work can draw on richer behavioral signals and broader datasets.

\section{Related Work}
Prior work has evaluated AI tools such as Copilot and Codex in general-purpose repositories, showing improvements in correctness \cite{chen2025codemeincreasingai} and how agent-driven refactoring impacts code quality \cite{horikawa2025agenticrefactoringempiricalstudy}. Studies have also explored review process of AI-authored PRs \cite{watanabe2025useagenticcodingempirical} and how developers integrate LLMs as features in mobile apps \cite{hau2025llmsmobileappspractices}. More recent mobile-focused research finds that while LLMs can generate simple code, they struggle with real-world complexity, as shown by APPFORGE’s end-to-end Android failures \cite{ran2025appforgeassistantindependentdeveloper}, studies of unreliable Flutter code \cite{10436306}, and case evidence that AI assistance accelerates tasks yet still demands substantial human correction \cite{10398656}. Industry trends echo this limited adoption, with 52\% of developers reporting little or no use of agentic tools \cite{stackoverflow2025devsurvey}. We extend this literature by studying 2,901 AI-authored PRs from 193 Android and iOS open-source projects, providing platform-level comparisons and the first task-category trend analysis of AI-generated work in mobile development.


\section{Discussion}

Our results reveal distinct platform-level differences in how AI agents contribute to mobile development. As shown in Section~\ref{sec:results}, Android exhibits higher PR acceptance rates and greater variation across agents and categories, whereas iOS behaves more uniformly. The agent- and category-level analysis further show that \textit{Codex} performs strongest overall, routine changes resolve quickly, and structural updates require substantially longer review time.

For a rough idea of the reasons behind PRs rejection on both platforms, we manually inspect a random sample of 10 rejected PRs which revealed additional potential causes: rebases, request-limit errors, agent-side execution failures, failing CI pipelines, missing reviewer feedback, and auto-closures due to inactivity. These observations clarify integration challenges that arise independently of PR task category or agent choice.

\textbf{Implications.} The combined results highlight where agent-generated changes align well with existing review practices and where they encounter delays. These insights support clearer expectations around acceptance likelihood and review effort across mobile platforms, agents, and task categories.

\textbf{Future Work.} Future analysis can extend this study to cross-platform frameworks (e.g., Flutter, React Native), incorporate additional collaboration signals such as review comments and future defects, and examine long-term trends as agent models evolve. Evaluating private industrial repositories can further test whether the patterns identified here generalize beyond open-source settings.

\section{Ethical Considerations}
Our study analyzes AI-authored pull requests from the publicly available AIDev dataset and does not involve interaction with human subjects. We cite the dataset authors throughout the paper and report results only at an aggregate level to reduce potential privacy or reputation risks.

\section{Conclusion}

We present an empirical study of $2,901$ AI-authored pull requests across $193$ Android and iOS open-source GitHub repositories from the AIDev dataset. Android projects exhibit higher PR acceptance rates and greater agent and category-level variation than iOS. Task categories such as \textit{}{localization}, \textit{fix}, and \textit{ui} PRs show the highest acceptance rate, while non-functional PRs remain slower to resolve on both mobile platforms. PR resolution times improve through mid-2025 on Android, while remain uniform but faster on iOS.
Our results provide baseline evidence of how current AI coding agents interact with mobile platforms, highlighting reliable performance on routine tasks such as \textit{fix} and persistent limitations on more structural modifications like \textit{refactor}. Our findings offer a basis for future evaluations of agent behavior and for developing platform-aware techniques that better support agentic mobile development.

\bibliographystyle{ACM-Reference-Format}
\bibliography{references}

\end{document}